# PERCCOM: A Master Program in Pervasive Computing and COMmunications for Sustainable Development

Jari Porras
Lappeenranta University of Technology
117 PUBLICATIONS   211 CITATIONS
SEE PROFILE

Eric Rondeau
University of Lorraine
139 PUBLICATIONS   457 CITATIONS
SEE PROFILE

Karl Andersson
Luleå University of Technology
56 PUBLICATIONS   184 CITATIONS
SEE PROFILE

# PERCCOM: A Master Program in Pervasive Computing and COMmunications for Sustainable Development

*Jari Porras, Ahmed Seffah*
Lappeenranta University of Technology
Lappeenranta, Finland
Jari.Porras@lut.fi, Ahmed.Seffah@lut.fi

*Karl Andersson*
Lulea University of Technology
Lulea, Sweden
karl.andersson@ltu.se

*Eric Rondeau*
University of Lorraine
Nancy, France
eric.rondeau@univ-lorraine.fr

*Alexandra Klimova*
ITMO University
St. Petersburg, Russia
alexandra.kgsu@gmail.com

*Abstract*— **This paper presents the Erasmus Mundus Joint Master Degree in Pervasive Computing and Communications for Sustainable Development (PERCCOM). This program brings together 11 academic partners and 8 industry partners to combine advanced Information and Communication Technologies (ICT) with environmental awareness to enable world-class education and unique competences for ICT professionals who can build cleaner, greener, more resource and energy efficient cyber-physical systems. First, this paper describes the rationale and motivations for ICT education for sustainability challenges. It then details the structure and contents of the programs including the courses offered at the three teaching locations (Nancy France, Lappeenranta Finland, and Lulea Sweden). The ways that the program has been running as well as students selection, their thesis works, involvement of industry, are also discussed. The program was built and managed using a solid academic standards and strategies student-centered learning.**

## I. Sustainability Development and ICT: Education Needs and Challenges

Sustainability and sustainable development have been defined in different ways by diverse communities. Perhaps the most know definition for sustainability is given in the so-called Bruntland report [1] to refer "*a development that meets the needs of the present without compromising the ability of future generations to meet their own needs*". The key issue here is the limits we should consider while developing our business, technology, society, etc. This old definition is later supported by the UN Secretary-General Ban Ki-moon with his statement: "*We hold the future in our hands, together, we must ensure that our grandchildren will not have to ask why we failed to do the right thing, and let them suffer the consequences.*" (Sustainable development at http://www.unesco.org, 2007).

Sustainable development is usually divided into three perspectives; economical, ecological and social [1].

Depending on the perspective taken to the sustainable development different aspects are emphasized. If looking the natural ecosystems and their limits the sustainability takes the ecological perspective. Sustainability is defined in that context as the environment's capability to sustain the population and to adjust to changes. Business sector looks at the sustainability from the economical perspective while following the same principles; environment needs to be able to sustain the business population. While businesses change their visions the business sector changes, some companies flourish and other may suffer. The social perspective of the sustainability has not been analyzed and discussed as much as the two other perspectives although people will, in the end, be in the key role of achieving the sustainability. Too often people are more linked to the economical sector than ecological and decisions are mostly made with economy in mind As Joseph Tainter states in his article [2] "*People sustain what they value, which can only be derived from what they know*". In order to achieve sustainability within ecological and economical limits, we need to change the people and their habits.

In addition to the three most cited perspectives, technical and individual perspectives have recently been added to the definition [3; 4]. Penzenstadler defines in [3] the technical perspective as "*longevity of systems and infrastructure and their adequate evolution with changing surrounding conditions*". In this definition technical perspective could be seen mainly as a target for sustainability rather than a possible tool. This type of an approach is commonly referred as Green IT, an approach that intends to improve the efficiency of the IT throughout its lifecycle. Various Green IT -type of approaches have been presented to get some steps closer to sustainable future. If considering also the tool aspect the technology could have a more important role in achieving sustainability. Bill Tomlinson in his book Greening through IT [5] emphasizes the role of IT/ICT in achieving sustainability in other domains as well. Various reports, including Simon Mingay's Gartner reports "Green IT: The new Industry shock wave" in 2007 and "The Impact of Recession on Green IT: Survey Results" in 2009, OECD report "OECD Information Technology Outlook 2010", Ken McGee's report "The 2011 Gartner Scenario: Current States and Future Directions of the IT Industry" and many others expressed the need for ICT to step forward and lead the transition to an energy-efficient and low-carbon economy.

Individual perspective is closely related to the social perspective but considers the impact of sustainability for an individual point of view [4] rather than society point of view. In many cases the individual make the personal decision for or against sustainability. In order to make a change and to change society and people we need to educate them. Like Stephen Downes state it "*We need to move beyond the idea that an education is something provided for us, and toward the idea that an education is something that we create for ourselves*" (http://www.huffingtonpost.com, 2010).

Within the context of Computer Science and Software Engineering, sustainability is a rather new perspective. While first sustainable ICT articles are from the beginning of the millennium, the focus moved on Software Engineering just few years back [10]. It has been also reported that traditional Software Engineering has not fully supported sustainability [3]. Moreover, there is little guidance on how Software Engineering can contribute to improving the sustainability of the systems under development [11]. We software engineers approach specific topics that have to do with sustainability in our discipline, for example, green IT, efficient algorithms, smart grids, agile practices and knowledge management, but we lack a common understanding of the concept of sustainability and if and how it can be applied or integrated to software engineering. Few curricula proposals and programs are emerging here and there [12; 13; 14] proposing some courses or programs in software sustainability or green software engineering. However, the most recent releases of the IEEE/ACM Curriculum Guidelines for Undergraduate Degree Programs in Software Engineering [15] or Computer Science [16] do not provide any indication to those topics. Sustainability within and through Computer Science and Software Engineering needs to be addressed better in near future.

In this paper we intend to present an existing approach, PERCCOM programme, to educate students on sustainable development in IT domain. We present the objectives as well as contents and some results of the programme and analyze them against some requirements presented in literature. More over we present the lessons learned after 3 years of running this programme.

## II. PERCCOM: OBJECTIVES AND STRUCTURE

PERCCOM programme is a European Erasmus Mundus Joint Master Programme in Pervasive Computing and Communications for Sustainable Development. This programme is an effort of 18 academic and industrial partners to fill the gap in sustainable (ICT) development education. PERCCOM programme was created during 2011-2013 to address the following specific challenges:

- To understand the emerging sustainability challenges within society and businesses and to transfer them into educational solutions with ICT as a key element,

- To combine the strengths, competences and experience of experts in different ICT perspectives (e.g. elements of systems as well as whole systems, hardware as well as software, communication and computation, a single phase of an element as well as the whole lifecycle) and thereby develop a common platform of competence within the guidelines of the Bologna process,

- To propose the new International Master degree with no currently available match at international level filling the gap between ICT skills and environmental considerations,

- To attract highly motivated international students to take this challenge and to create new solutions,

- To provide the prospective students with knowledge, skills and finally competencies in sustainability and ICT to enable a true impact on ecological, economic and social aspects of sustainability, and

- Finally to fulfill the needs presented by those various reports on ICT's role as a solution or a part of it.

Figure 1 portrays the sustainability perspectives addressed in the PERCCOM programme. The main skills required and emphasized will be ICT based. After graduating the students will have the fundamental skills on computer networking, software development as well as ability to combine and utilize these under complex systems. Thus, the programme will not be a traditional Software Engineering or Computer Science programme, but a mixture of them.

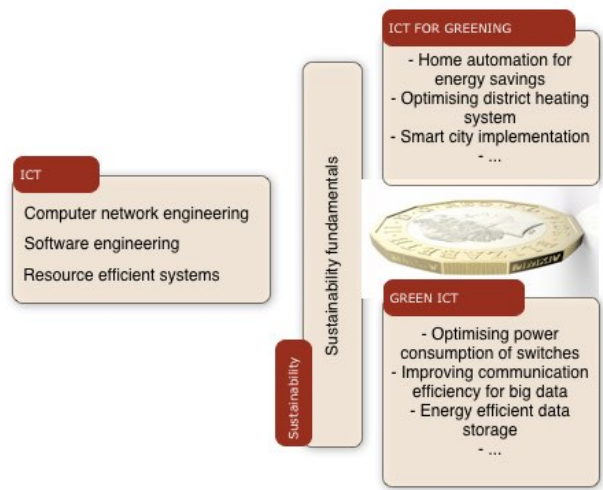

Figure 1. Skills emphasized in PERCCOM.

Each and every ICT course is tied to sustainability fundamentals and skills. This gives the students a possibility to study how ICT technology can be made more efficient (Green ICT) or how ICT can be used to make various application areas more efficient (ICT for Greening). Although Green ICT and ICT for greening can be seen different sides of a coin, they can easily co-exist and co-evolve. Figure 1 shows few research (thesis) topics of the students under these perspectives. These research works have been conducted and reported by the first set of students of the programme in June 2015. The thesis topics under Green ICT are focusing on improving the efficiency of the technology, (e.g. energy efficiency of network switches) while thesis topic under ICT for greening use ICT for improving sustainability of different domains, e.g. energy savings at home or whole districts.

The general structure of the PERCCOM programme is presented in Figure 2. The PERCCOM master's programme lasts for two years and is divided into 4 semesters. Three semesters are used for the education while the last semester is reserved for the master's thesis. Each semester consist of ICT courses (including industry seminars) with sustainability flavor, student project focused on master's thesis project, and cultural courses. Courses on local language and culture, as well as industrial seminars organized with international enterprises are important due to the growing demand for graduates capable to work in international environment. The thesis topic and project follows the students throughout the programme. After selecting their topics in first semester the subsequent project courses in different semesters take the students deeper into their topic, research and research methods. Students also have a possibility to consider in all other courses their own thesis topics and reflect the learning outcomes towards that. In the following the semesters and courses of the programme within the semester are further described.

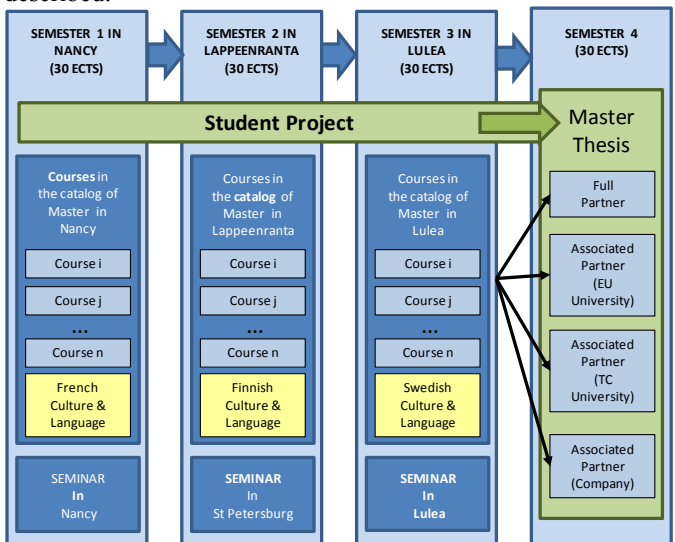

Figure 2. An overview of structure and contents of PERCCOM programme.

*A. Semester 1 - Sustainable Computer Network Engineering*

Each selected cohort of students will start its studies in Nancy, France, with the computer network engineering focused semester. The objective of this semester is to provide students with fundamental competences in computer networks and systems engineering in a sustainable way.

- Communication protocols (6 ECTS) – ICT and Sustainability

  This lecture focuses on networks fundamentals. It describes the standardization process, the major network architectures, and the main concepts such as switching, encapsulation, errors recovery, connection principles, etc. Classical protocols are presented and detailed. Sometimes, their drawbacks in terms of energy consumption, resources utilization, etc. are highlighted, and some environmental-aware solutions are described.

- Quality of Sustainable Service (6 ECTS) – ICT and Sustainability

  The course provides knowledge on Quality of Service (QoS), Quality of Experience (QoE) and Quality in Sustainability (QiS) in communication network domain. It gives theoretical and practical knowledge on how to maintain or improve QoS/QiS level at lower layers of the network in terms of delay, reliability, security, energy consumption, recycling, etc.

- Automatic Control for Sustainable Development (3 ECTS) – ICT and Sustainability

  The Research Center for Automatic Control in Nancy (CRAN) is participating and managing many research projects in the environmental domain. The objective of this course is to present to students an umbrella of results on how using automatic control in the context of sustainable development.

- Systems Engineering (3 ECTS) – ICT, Sustainability and Green ICT

  Model-Based Systems-Engineering (MBSE) is presented as a key iterative, collaborative and multidisciplinary approach to define, to develop and to deploy systems in general. MBSE is then applied to ICT domain in order to enhance a better interoperation between system components, humans as well as technologies, and then to contribute to meet eco-efficiency for C2C (Cradle to Cradle) issues beyond conventional Business and PLM efficiency.

- Sustainable development & circular economy (3 ECTS) - Sustainability

  This course provides an introduction to the general and sectorial regulations of the European Environmental legislation with respect to selected matters of the economic legislation. Competences, principles and instruments based on the Lisbon treaties, problems of horizontal secondary legislation especially concerning the access to information, process participation, remedy, EIA and environmental liability as well as selected topics of the sectorial plant and product related environmental legislation.

- Specification definition of Master thesis project (3 ECTS) – ICT and Sustainability

  The students start their master thesis from semester 1. All the topics are related to Green IT and sustainable development.

- French Culture and Language (3 ECTS) – Sustainability (social aspects)

The students start the semester 1 with intensive courses in French to help them in their daily-life in France. Different cultural events are also organized enabling students to discover French art, French history, French food culture, etc.

After the semester 1 the students are expected in sense of sustainable development to be able
1) to utilize (new) green ICT metrics to evaluate a level of sustainability of ICT systems and to formalise new green SLA (Service Level Agreement) between ICT companies and their customers. Original approach based on Biomimicry is explained to students to show how to imitate the nature to develop ICT solutions in the context of sustainable development;
2) to utilize (new) standards and tools for measuring energy consumed by ICT systems. For example, the students use Cisco Energywise protocol to monitor energy consumed by network architectures, GreenSpector tool to measure energy consumed by software. It conducts the students to elaborate original strategies to efficiently design networked software applications;
3) to be systemic when developing Green ICT applications in considering all the stakeholder requirements, the whole life cycle of ICT systems and all the facets of sustainable development (and not only energy). For that, study cases are analysed using system engineering methods to eco-design ICT systems;
4) to have expertise in scientific methods for creating new mathematical models on energy consumed by ICT systems. For example, the students use design of experiment for modelling energy consumed by switches.

*B. Semester 2 – Smart software and services*

The second semester is implemented by LUT in Lappeenranta Finland and ITMO in St. Petersburg, Russia and will concentrate on software and service issues as software will be in key role when ICT is used for achieving sustainability in different application domains. The objective of this semester is to provide students with Software Engineering fundamentals and sustainability perspective to software systems.
- Green IT and Sustainable Computing (5 ECTS) – Sustainability and ICT for Greening
  This course presents the ICT for greening approach for the students. This is the only course in second semester that completely focuses on sustainability issues. The sustainability issues are tackled with changing set of books and scientific articles focusing on various aspects of software (ICT) solutions for sustainability. This course follows the flipped-classroom principles and aims to increase the critical thinking and argumentation skills as well as reflection competencies.
- Architecture in Systems and Software Development (7 ECTS) - ICT
  This course focuses on the software development fundamentals, especially the architecture aspects. The student understands the role of architecture in the development of software and information systems and has the basic skills of how to design and describe architecture. Sustainable software development aspects will be emphasized as factors affecting the software development. This course follows traditional lectures, exercises, practical work and exam approach.
- Software Systems as a Service: Technology and Engineering (5 ECTS) – ICT and ICT for Greening
  This course emphasizes the ongoing shift in software and application development though service orientation. The sustainability aspect of the service is emphasized and considered like any other quality attribute of the software. Students are encouraged in their projects to focus on sustainability issues. This course follows a project-based approach emphasizing practical solutions for sustainability.
- Code camp on Platform Based Development (4 ECTS) – ICT and ICT for Greening
  Code camp is a highly practically oriented course on software development. Although this course is provided 3-4 times per year with variable content, the topic for PERCCOM students is always related to the sustainable development. During spring 2015 the PERCCOM students implemented various open data applications, demonstrating the usability of different open data sources.
- Transformation of a modern industrial society: The Finnish Model (2 ECTS) – Sustainability (social aspects)
  This course represents the cultural content of the second semester. It was seen more important to teach the students the culture of Finland and Finnish work markets than language as such. This gives the students a cultural perspective to the social sustainability in Finland.
- Towards Semester 3 (1 ECTS) – Sustainability and ICT
  Towards semester 3 course has two focal points. First it progresses the research of each PERCCOM student by teaching students how to make appropriate literature reviews for their thesis topics and how to tie their topic to the concept of sustainability. Second, it prepares the students for the third semester that will start after a summer break.
- ITMO - Seminar 1 & 2 (6 ECTS) – Green ICT
  The industry seminars in second semester are arranged together with the St. Petersburg National Research University of Information Technologies, Mechanics and Optics (ITMO University) in St. Petersburg. These seminars have emphasized various cloud services and technologies needed by them as well as cultural issues in Russia.

After the semester 2 the students are expected in sense of sustainable development to
1) understand the meaning of software for sustainable development. The software solutions can be applied

in various application fields from home automation to smart transportation.
2) understand the social (human) perspective and its impact on sustainable development. Software solutions give us tools for affecting on the human behaviour.
3) be able to implement software solutions for various problems by considering the metrics and methods for sustainable development.

*C. Semester 3 - Resource efficient pervasive computing systems and communication*

The third semester focuses on wireless networking and systems perspective as well as combining networking and software into efficient pervasive systems.

- Network programming and distributed applications (7.5 ECTS) – ICT, ICT for Greening

   The learning outcome of this course is the scientific foundation of network programming and distributed applications including security considerations and the proven experience programmers in this field of Computer Science. Furthermore, students are given the capacity for carrying out teamwork and collaboration with various constellations, both in groups where the students choose whom to work with and in groups put together by others. The goal is also that students can create, analyze and critically evaluate various technical solutions in terms of the design and implementation of communicating computer programs and to show insight in research and development by understanding limitations and possibilities. Also, they should be able to plan and use appropriate methods to undertake advanced programming tasks within predetermined parameters and show the ability to identify knowledge gaps and bridging these gaps by gaining new knowledge. Last, but not least, they should gain the ability to understand, interpret and present scientific publications in the area. Teaching methods include traditional lectures by academic staff members, guest lectures by industry representatives, lab assignments, and seminar presentations by students. The course is finished with a home exam.

- Wireless sensor networks/ Wireless Mobile Networks (7.5 ECTS) – ICT and Green ICT

   The learning outcome of this course covers details of current radio transmission technologies on the physical layer. Students study how to compute the parameters of the radio transmission zone, to describe the problem of hidden/exposed terminal problem and explain the difference between the existing MAC protocols for wireless and wired networks, to explain the channel capture effect on MAC and Transport layers, and to describe the concept of proactive and reactive routing protocols. Furthermore, they should be able to describe the concept of geographical and content based routing, and problems of transport layer protocol over multihop wireless networks and present existing solutions. Last, but not least, they should be able to describe a simulation scenario and plan the simulation experiments like TCP and UDP protocols over different types of ad hoc routing protocols in a network simulator and measure throughput, packet loss rate performance characteristics. They should also be able to explain the functionality of LTE, WiMax, Zegbee and Bluetooth network architectures. Teaching methods include traditional lectures by academic staff members, guest lectures by industry representatives, lab assignments, and seminar presentations by students. The course takes advantage continual examination with smaller exams given throughout the course.

- Multimedia systems (7.5 ECTS) – ICT and ICT for Greening

   The learning outcome of this course covers knowledge about distributed systems for global distribution of real-time media for human communication. It trains students in their capacity for carrying out teamwork and collaboration with various constellations, both in groups where the students choose whom to work with and in groups put together by others. Furthermore, the goal is to have students capable of creating, analyzing and critically evaluating various technical solutions in terms of the design and implementation of multimedia systems and to show insight in research and development by understanding limitations and possibilities. Last, but not least, tis course teaches students to plan and use appropriate methods to undertake advanced programming tasks within predetermined parameters and show the ability to identify knowledge gaps and bridging these gaps by gaining new knowledge. Teaching methods include traditional lectures by academic staff members, guest lectures by industry representatives, lab assignments, and seminar presentations by students. The course is finished with a home exam.

- Special Studies in Pervasive and Mobile Computing (Project) (3 ECTS) – ICT and Sustainability

   The learning outcome of this course covers practical aspects of group project where the assignment is a current and interesting problem in the area of distributed computing systems for sustainable development. Students are introduced in Agile development and expected to use the knowledge they have acquired in previous courses and search and study literature to solve the task. So far, students within PERCCOM have been participating in the Green Code Lab Challenge collaborating within teams composed of members both in Luleå, Sweden, and Nancy, France. Teaching methods include traditional lectures by academic staff members, guest lectures by industry representatives, and the above-mentioned project assignment.

- Swedish for Beginners AI:1a - (3 ECTS) – Sustainability (cultural aspects)

   The students start semester 3 with an intensive courses in Swedish with the learning outcome to help them in their daily-life in Sweden so that the student can present themselves and their background in Swedish, be able to read and understand simple Swedish texts, and be able to make use of basic knowledge about the structure of the Swedish language. Teaching methods include traditional lectures by academic staff members. In addition, cultural visits are performed during this course. The course is finished with a traditional written exam.

- Seminar: (1.5 ECTS) – ICT and Sustainability

  The learning outcome of this course is the aggregated content of a set of guest lectures by invited academic staff members and senior executives within the IT industry. At the end, students present the status of their own thesis work and submit summaries of the seminars.

*D. Semester 4 - Master's thesis*

Although the fourth semester is solely reserved for the Master's thesis, the selected students start their thesis work already during the first semester. As each student has a defined thesis from the first semester on and they will have a possibility to work on their thesis topic on all semesters, the PERCCOM students are expected to show all the expected skills of PERCCOM programme, especially to combine sustainability elements into their thesis topic domains. The following topics represent examples from the thesis works of the first cohort students graduated in September 2015. These topics emphasize different sustainability perspectives within PERCCOM programme.

- *Energy consumption of applications on mobile phones*

  This work focuses on ICT, Green ICT and sustainability issues. Sustainability is emphasizing the energy, i.e. ecological issues although otherwise the work is tightly connected to ICT, especially software development issues.

- *Green aspects study in game development*

  If the previous work combined sustainability to ICT with very tight and practical way, this topic focuses more on software development processes and sustainability knowledge within gaming industry. As such this work combines sustainability to ICT and does not so much contribute either to Green ICT or ICT for greening perspectives.

- *Green service level agreement under sustainability lens in IT industry [17]*

  Similar perspective is taken in this topic. The objective of the work is to define the ways how sustainability issues could be brought into service level agreements. As such this work also emphasizes the connection between ICT and sustainability knowledge.

- *Analyzing and computing the sustainable gains of building automation [18]*

  This work represents the ICT for greening approach. The technical fundamentals lie in home automation systems and the sustainability is considered both from economical and ecological perspectives.

- *Analyzing the power consumption behavior of Ethernet switch using Design of Experiment [19]*

  This work represents quite pure green ICT topic. The objective of the work is to minimize the energy usage or Ethernet switches and as such contribute to the ecological sustainability.

The Master's thesis topics show wide variety of research domains the PERCCOM programme is dealing with. As these were the first topics given for the students they represents more of the perspectives of different participants rather than the combined understanding of the PERCCOM programme.

*E. Intended learning outcomes and markets for the graduates*

PERCCOM programme represents a combination of Computer Science and Software Engineering disciplines. The programme is aiming at providing the students a sustainable mindset for systems development regardless of the sustainability perspective (ecological, economical, social). Graduates of the programme will be capable to design, develop, deploy and maintain both pervasive computing systems and communication architectures with sustainable orientation. The focus on sustainability perspective does not diminish the role of core ICT skills and competences. The graduates could employ themselves to almost any networking or software related job a traditional software engineer or computer scientist would. The industrial program partners like Orange, Cisco, Ericsson AB, Facebook, represent international environments that will benefit of the programme graduates.

III. REFLECTING PERCCOM TOWARDS EXISTING KNOWLEDGE AND LESSONS LEARNED AFTER FIRST STUDENT COHORT

Although sustainability aspects in education, especially in ICT related education, have not been emphasized a lot in the past, there exist few articles that can be used for evaluating the contents and structure of the PERCCOM programme. The article of Samuel Mann et al. [20] presents three approaches (centralized, distributed and hybrid) of integrating sustainability issues within the curriculum and a framework for assessing the integration of sustainability issues. Cathy Rusinko presents a very similar idea for sustainability integration [12] by looking at the scope and structures of the sustainability programme. Figure 3 presents this model. Another similar concept is presented in [13].

|  | SHE delivery | |
|---|---|---|
|  | Existing structures | New structures |
| Narrow (discipline-specific) | I. Integrate into existing course(s) minor(s), major(s), or programs(s) | II. Create new, discipline-specific sustainability course(s), minor(s), major(s) or programs(s) |
| Broad (cross-disciplinary) | III. Integrate into common core requirements | IV. Create new, cross-disciplinary sustainability course(s), minor(s), major(s), or programs(s) |

(SHE focus)

Figure 3. Integrating sustainability into higher education [12].

None of the models presented in [12; 13; 20] perfectly match to the situation of PERCCOM programme. The programme is intentionally implemented as an international programme combining different perspectives of ICT education. Although all hosting partners of PERCCOM are in ICT field the outcome almost reaches multidisciplinary nature due to wide variety of ICT perspectives. PERCCOM programme as such is a new structure that would probably not have happened in a single university. Each university has learned from others and got advantage of their perspectives. Each hosting university could also have used the PERCCOM programme as a part of their own structures.

For example in Lappeenranta University of Technology PERCCOM programme itself is a new structure and extends the perspective of the ICT discipline. Few completely new courses have been implemented due to PERCCOM programme and in others the sustainability aspects have been applied. In addition the Lappeenranta University of Technology has implemented sustainability minor for the students, in which PERCCOM knowledge and courses play important part. So from the Lappeenranta University of Technology perspective PERCCOM can be seen as a part of quadrants I, II and III. Other hosting partners may see the programme in a very different or similar manner depending how the universities want to utilize the sustainability knowledge.

The implementation of courses within each semester depends heavily on how the hosting partners have approached the sustainability integration. Even a priori, all these courses of the programme look very similar to courses in any CS or Software Engineering Software Engineering program. However, when the PERCCOM programme was created it was required that sustainability has to be addressed as a central topic in all the courses. Nobody at that point considered that the perspective to the sustainability might differ. With the first set of students the consortium has learned the different perspectives towards the integration of ICT and sustainability and the programme has evolved and is continuously evolving. For example, the course on Software Systems as a Service: Technology and Engineering in semester 2 has evolved from a rather technically oriented course to a course that addresses the successful and failures stories from industry on the integration of sustainability in the service orientation approach. A large team-oriented project on service systems for sustainability innovation is also part of the course. Overall, in this course and in many others sustainability has evolved in various levels. In this course particularly development has focused on large-scale understanding of sustainability (i.e. sustainability is not only covered from technical and ecological perspectives but also from other perspectives (economical and social)):

- Similar to security and other software quality attributes, sustainability is defined as a key quality attribute of a service system
- Students are encouraged to consider projects related to the re-engineering of existing software systems and/or the development of innovative services to support sustainability development including the management of natural resources consumption as well as the ways software services can make citizens more aware about their impacts on the environment.

The students (c. 20) have been selected from a large international pool of students (400+) with various cultural and educational backgrounds. In particular, three programme cohorts are represented by more than 30 nationalities from four continents with educational background mostly in Software Engineering, Computer Science, Computer Engineering, Electrical Engineering and Automation Engineering, emphasizing the various skills, e.g. software development, programming, networking, and automation. This is one of the key ingredients that contribute to the overall success of the program. It also helps to developing somehow the societal aspects of sustainability. The awareness of sustainability and green technologies among students is similar. This is in line with the studies from different parts of the world that show many similarities in the views of youth about sustainability, regardless of ethnicity, race, sex, or geographic location. These common perceptions were the starting point of all the discussions on the effort around their current and future roles and responsibilities developing and managing sustainability and green software. The different sustainability perspectives of hosting partners, the cultural studies included into the PERCCOM program and the different cultural backgrounds provide a fruitful opportunity for sustainability education. The cultural perspective enables the emphasis of social perspective of sustainability in addition to the ecological and economical perspectives.

For example, during the semester at Lappeenranta University of Technology students were asked too investigate how to use software systems to minimize the environmental consequences caused by humans. The key objectives are:

- Empowering people means using ICT to raise people's awareness of the environmental impact of their actions and to channel their behavior in a more environmentally-friendly direction
- Extending natural resources involve reducing the use of diverse environmentally unsustainable resources through ICT-based solutions
- Optimizing systems refers to minimizing the environmental load of diverse systems by optimizing their operation

Students have been encouraged to participate in the green.citizen@ICT project [21] while developing and validating new services focusing on citizen engagement to reduce natural resources consumption and, on ways to increase awareness in daily activities such as car driving, house activities and, shopping.

The graduates of the first cohort have demonstrated the multiplicity of future possibilities. Out of 17 graduates
- 9 continue their studies in various doctoral programmes. 5 of these topics are closely related to green technologies and sustainable development. Other topics are either software or networking focused.
- 4 have employed in software or network development positions.

- 2 continue their work in various management positions in which they have possibility to consider sustainable development.
- 2 have moved to academic positions.

## IV. Discussion and education foundations of the program

As we already highlighted, it is not an easy task to build and run such programs in the absence of standards and clear guidelines. Traditional education standards and curriculum development tools such ACM/IEEE curriculum guidelines for computer science and software engineering education as well as SWEBOK (Software Engineering Body of Knowledge) did not provide any indication on sustainability, green IT or green software engineering. Research and investigations are needed to building a stranded curriculum as universities are running programs and providing feedback to the community. This paper is a contribution to these efforts.

Sustainability in academic research and education is rather young area as it really started to evolve in the beginning of this millennium and as such the education of sustainable ICT is also evolving. At the starting time of PERCCOM programme there were no programmes focusing on ICT and sustainability aspects. Although some research groups have been active in publishing articles concerning ICT and sustainable development the research results have not got into actual education. PERCCOM program, as the first of the kind, is by for not perfect. Each hosting partner had in the beginning its own perspective to sustainability and the implementation of sustainability within courses varied. With the first set of students the PERCCOM consortium has evolved and the PERCCOM program is approaching something that could be multiplied to other locations and other networks. This paper has presented few lessons learned from the structure and contents of the whole programme but still miss deeper analysis of single courses. This is to be done in future papers. What is clear so far is that the highly motivated and talented students and well collaborating universities can create sustainability programme with almost multidisciplinary support. In future the sustainable ICT programmes could be even further divided into programmes focusing on sustainable networking, software for sustainable development, etc.

## Acknowledgements

The PERCCOM program is funded by a grant from the European Union ERASMUS Mundus program. The authors would like to acknowledge the European Union. We also thank all the PERCCOM students from around the World.